\documentclass[aps,pre,reprint,superscriptaddress,raggedbottom]{revtex4-2} 

\usepackage[svgnames, table]{xcolor} 

\usepackage{graphicx} 

\usepackage{amsmath,amssymb,amsfonts} 
\usepackage{bm}

\definecolor{linkColor}{RGB}{0,70,120}
\usepackage[colorlinks=true, allcolors=linkColor,pdfborder={0 0 0},pdfencoding = auto]{hyperref}

\usepackage[normalem]{ulem}

\definecolor{fb_color}{rgb}{0.8,0.5,0.}

\definecolor{mo_color}{rgb}{0.8,0.0,0.6}

\definecolor{new_color}{rgb}{0.,0.4,0.1}

\begin{document}

\title{Active Solids: Topological Defect Self-Propulsion Without Flow}
\author{Fridtjof Brauns} 
\email{fbrauns@kitp.ucsb.edu}
\affiliation{Kavli Institute for Theoretical Physics, University of California Santa Barbara,
Santa Barbara, CA 93106, USA}
\author{Myles O'Leary}
\email{mo6344@princeton.edu}
\affiliation{Department of Physics, Princeton University, Princeton, NJ 08544, USA}
\author{Arthur Hernandez}
\email{hernandez@lorentz.leidenuniv.nl}
\affiliation{Instituut-Lorentz, Leiden University, P.O. Box 9506, 2300 RA Leiden, The Netherlands}
\author{Mark J. Bowick}
\email{bowick@kitp.ucsb.edu}
\affiliation{Kavli Institute for Theoretical Physics, University of California Santa Barbara,
Santa Barbara, CA 93106, USA}
\affiliation{French American Center for Theoretical Science, CNRS, KITP, Santa Barbara}
\author{M.\ Cristina Marchetti}
\email{cmarchetti@ucsb.edu}
\affiliation{Department of Physics, University of California Santa Barbara, Santa Barbara, CA 93106, USA}

\begin{abstract}
    The self-propulsion of $+1/2$ topological defects is a hallmark of active nematic \emph{fluids}, where the defects are advected by the flow field they themselves generate. In this paper we propose a minimal model for defect self-propulsion in a nematic active \emph{solid}: a linear elastic medium with an embedded nematic texture that generates active stress and associated elastic strains. We show that such coupling gives rise to self-propelled $+1/2$ defects that move relative to the elastic medium by local remodeling of the nematic texture without advection. This mechanism is fundamentally different from the fluid case and can lead to unbinding of defect pairs and stabilization of $+1$ defects. Our findings might help explain how orientational order, of, for example, muscle fibers, is reconfigured during morphogenesis in solid-like tissues. The proposed mechanism may, for instance, control motility and merging of $+1/2$ defects, which play a crucial role in setting up the body axis during \textit{Hydra} regeneration.
\end{abstract}

\maketitle

Topological defects
are found in many natural and synthetic systems. They are the organizing centers of orientational order fields and have been implicated in morphogenesis \emph{in vivo} \cite{Maroudas-Sacks.etal2021,Maroudas-Sacks.etal2024confinement,Ravichandran.etal2025,Sinigaglia.etal2020} and \emph{in vitro} \cite{Guillamat.etal2022,Dessalles.etal2025,Shen.etal2025}, as well as in the organization of tissues~\cite{Kawaguchi.etal2017,saw2017topological} and bacterial colonies~\cite{Copenhagen.etal2021}.
In synthetic nematic elastomers, topological defects can be used to control shape morphing. In these materials internal forces can be generated through chemical or thermal activation of the elastomer, yielding nemato-elastic stresses that deform the system \cite{Warner.Terentjev2003,deHaan.etal2012,Warner2020}.
Theoretical studies of such shape-morphing solids have focused on quantifying the deformations induced by prescribed (``frozen-in'') nematic textures~\cite{Modes.etal2011,Mostajeran2015,Duffy.Biggins2020,Khoromskaia.Salbreux2023}.
On the other hand, dynamic reconfiguration of nematic textures through defect creation, motion and fusion has been shown to play an important role in morphogenesis, as studied for instance, during whole-body regeneration of the freshwater polyp \textit{Hydra}~\cite{Maroudas-Sacks.etal2021,Sinigaglia.etal2020,Maryshev.etal2020,Ravichandran.etal2024}.

Self-propelled defect dynamics has been studied extensively in active nematic liquid crystals~\cite{shankar2022topological}. In these fluids, active stresses generate flows which in turn drive defect motion through advection~\cite{giomi2013defect,pismen2013dynamics}.
By contrast, defects in solid-like epithelial tissues have been observed to move \emph{relative} to the surrounding elastic material as the nematic texture remodels locally~\cite{Maroudas-Sacks.etal2021,Maroudas-Sacks.etal2024confinement}. 
In such tissues, cells adhere to one another strongly and their rearrangements happen much slower than the remodeling of intracellular filaments \cite{Lopez-Gay.etal2020,Maroudas-Sacks.etal2021,Mongera.etal2023,Shen.etal2025}, calling for a description as active nematic solids rather than fluids.

Here, we investigate the \emph{dynamics} of topological defects in an active nematic elastomer, where forces are generated internally by active processes. We propose a new mechanism for defect motion through local melting of the nematic texture driven by strains induced by active stresses. This mechanism does not require flow of material, providing  a natural framework for understanding the observed restructuring of the nematic texture in \textit{Hydra}. 
We propose a minimal model for a linear elastic medium with an embedded nematic texture that generates active stress and in turn is coupled to elastic strain. We show that such coupling gives rise to self-propelled $+1/2$ defects that move relative to the elastic medium by local remodeling. If the self-propulsion is sufficiently large, $+1/2$ defects can escape the Coulomb-like attraction to a $-1/2$ defect. 
In confined systems where defects are enforced through boundary anchoring, the elasto-nematic coupling gives rise to rich dynamics, including orbiting $+1/2$ defects, and stabilized $+1$ defects that can spontaneously deform from aster/vortex configurations to spirals.

\textit{Minimal elasto-nematic sheet model.}
The main motivation for our model is epithelial tissue with nematically organized acto-myosin fibers, as found for instance in the ecto- and endoderm of \textit{Hydra} \cite{Philipp.etal2009,Aufschnaiter.etal2017,Maroudas-Sacks.etal2021}. 
We denote the displacement field by $u_i$ and the (linear) strain tensor as $u_{ij} = (\partial_i u_j + \partial_j u_i)/2$.
Nematic order is described in terms of the symmetric, traceless tensor order parameter $Q_{ij} = S (n_i n_j - \delta_{ij} / 2)$, where $S$ is the magnitude of order and the director $\mathbf{n} \equiv -\mathbf{n}$, $|\mathbf{n}| \equiv 1$, describes the orientation of the nematic.
The free energy includes the usual Landau--de Gennes term for the isotropic--nematic transition, Frank elastic coupling in the single elastic constant $K$ approximation, and linear elasticity parametrized by the shear modulus $\mu$ and the bulk modulus $B$:
\begin{multline} \label{eq:free-energy}
    \mathcal{F} =
    \frac{1}{2} \int\!  \mathrm{d}r^2 \, \Big[-\left(1 + \chi_B^{} u_{kk}\right) Q_{ij}Q_{ij}
    + (Q_{ij}Q_{ij})^2 \\
    + K (\nabla_i Q_{jk} )^2 
    + 2 \tilde{u}_{ij} (\mu \tilde{u}_{ij} - \chi Q_{ij}) + B u_{kk}^{2}\Big],
\end{multline}
where we have chosen coefficients such that $2 Q_{ij}Q_{ij} = S^2 = 1$ for the uniform equilibrium when $\chi = \chi_B^{} = 0$.
The coefficient $\chi$ accounts for strain alignment of nematic order along the direction of deviatoric strain $\tilde{u}_{ij}$, aligning them when $\chi > 0$ and anti-aligning them for $\chi < 0$.
The strain alignment term can be derived by linearizing the Warner--Terentjev model for nematic elastomers \cite{Warner.Terentjev2003, Lubensky.etal2002}. It leads to a soft mode since the elastomer can accommodate shear through reorientation of the nematic.
The coefficient $\chi_B^{}$ describes the coupling of the nematic order to isotropic strain $u_{kk}$. We call this ``strain ordering.'' For $\chi_B^{} > 0$, nematic order decreases (``melts'') in regions of compressive strain ($u_{kk} < 0$).
To our knowledge, this term has not been previously considered (with the exception of Ref.~\cite{Maitra.Ramaswamy2019} where only defect-free systems are studied), since most nematic elastomer theories assume incompressibility.
In the context of biological tissue, both strain alignment and strain ordering can result from feedback-mediated regulation of the cytoskeletal organization in response to strain
\cite{He.etal2015,Gupta.etal2015,Maroudas-Sacks.etal2024confinement}. In this case $\mathcal{F}$ should be understood as an \emph{effective} free energy. 

We consider adiabatic dynamics in a regime of quasi-static force balance where the nematic texture remodels on a time scale much slower than that of force equilibration through deformations.
This assumption is justified in a tissue where forces balance on short time scales (seconds to minutes) compared to the remodeling of the actomyosin fibers that form the nematic texture (hours). 
Force balance within the elastic sheet requires $\partial_i \sigma_{ij} = 0$, where the total stress $\sigma_{ij} = \sigma^\mathrm{el}_{ij} + \sigma^\mathrm{a}_{ij}$ includes the passive elastic stress $\sigma^\mathrm{el}_{ij} = \frac{\delta \mathcal{F}}{\delta u_{ij}}$ and active isotropic and deviatoric stresses
$\sigma^\mathrm{a}_{ij} = \frac{\alpha_{B}}{4} S^2 \delta_{ij} + \alpha Q_{ij}$.
Positive (negative) signs of $\alpha_{B}$ and $\alpha$ correspond to contractile (extensile) activity.
Together, this gives
\begin{equation}
    \sigma_{ij} = 2\mu u_{ij} + (B-\mu)u_{kk}\delta_{ij}
    + \frac{\tilde{\alpha}_{B}}{4} S^2\delta_{ij} + \tilde{\alpha} Q_{ij}\;,
\end{equation}
where we have introduced the effective active stress strengths $\tilde{\alpha}_{B} = \alpha_{B} - \chi_B^{}$ and $\tilde{\alpha} = \alpha - \chi$ that capture the total isotropic and deviatoric stresses exerted by the nematic \footnote{If strain alignment and strain ordering result from biological feedback loops rather than passive (thermodynamic) mechanisms, $\mathcal{F}$ is only an effective free energy and the alignment terms do not contribute to the stress $\chi$ and $\chi_B$ in the stress. In this case, $\alpha = \tilde{\alpha}$, $\alpha_B = \tilde{\alpha}_B$.}.
We neglect passive stresses due to the nematic molecular field which are higher order in gradients than the active nematic stresses. 
Throughout the manuscript, we set $K = \mu = B/2 = 1$. Unless specified otherwise, we set $\chi_B^{} = -2$, $\tilde{\alpha}_B = 0$.

The nematic texture evolves through relaxational dynamics to minimize the free energy, 
\begin{equation} \label{eq:nematic-remodeling}
  \gamma \partial_t Q_{ij} = - \frac{\delta \mathcal{F}}{\delta Q_{ij}}\,,
\end{equation}
where
$\gamma$ is the rotational viscosity. We non-dimensionalize time such that $\gamma = 1$.
We assume a stiff elastic material, such that displacements are small, allowing us to neglect convective and co-rotational terms.

The effective active stresses induce elastic stresses which can (partially) relax through remodeling of the nematic texture. This is akin to the soft elasticity of nematic elastomers that accommodate \emph{external} deformation by reorientation of the nematic \cite{Warner.Terentjev2003}.
Geometrically, the induced elastic stresses can be understood as a consequence of metric incompatibility (intrinsic curvature) as is briefly discussed in Appendix~\ref{app:metric-elasticity}.
As the texture remodels, the \emph{active} stresses will, however, lead to further incompatibility. In other words, activity makes the coupling between elastic strain and nematic order non-reciprocal, thus allowing for the emergence of dynamic steady states~\cite{you2020nonreciprocity,saha2020scalar,fruchart2021non}, as shown below.

\begin{figure}[t]
    \centering
    \includegraphics{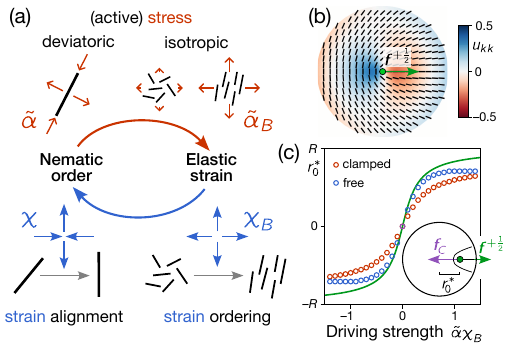}
    \caption{
    (a)~Illustration of the elasto-nematic couplings considered in our model. Black sticks illustrate nematic directors. Red and blue arrows show stress and strain respectively.  
    (b)~Gradient of isotropic strain across a $+1/2$ defect, causing the defect to self-propel (green arrow) through ``strain ordering'' in the tail (see Movie~1).
    (c)~Steady-state defect position $r_0^*$, where propulsion force $\boldsymbol{f}^{+\frac12}$ and Coulomb repulsion $\boldsymbol{f}_{C}$ due to anchoring at the boundary balance (see inset), as a function of active stress; numerics (circles) and analytic estimate (green line).
    (Parameters: $R=16, \tilde{\alpha} = 1, \chi = 0$.)
    }
    \label{fig:1}
\end{figure}

\textit{Self-propelled $+1/2$ defects.}
Only $+1/2$ defects have polar symmetry which is necessary for self-propulsion~\cite{narayan2007long}.
To investigate propulsion mechanisms in our model, consider a single $+1/2$ defect with its tail pointing along the direction $\hat{\mathbf{p}}$. Since we study an isolated defect, we can use rotational and translational invariance to set $\hat{\mathbf{p}} = \hat{\mathbf{x}}$, the positive $x$-direction and center the defect at the origin:
\begin{equation} \label{eq:Q+1/2}
    \mathbf{Q}_{+\frac12}(\mathbf{r}) = \frac12 S_{1/2}(r)
    \begin{pmatrix}
        \cos{\theta} & \sin{\theta} \\
        \sin{\theta} & -\cos{\theta}
    \end{pmatrix}\;,
\end{equation}
where $(r, \theta)$ denote polar coordinates. Equation~\eqref{eq:Q+1/2} describes the ``unperturbed'' equilibrium texture of a $+1/2$ defect with $\chi = \chi_B^{} = 0$.
The magnitude of order $S_{1/2}(r)$ vanishes at the defect core $r = 0$ and is $1$ for $r \gg \sqrt{K}$ with a transition zone (core size) of width $\sim \sqrt{K}$ \footnote{An approximate solution for $S_{1/2}(r)$ can be found using the Pad`e approximant \cite{shankar2018defect} (see SI)}.

Let us first consider the case of purely deviatoric induced stress ($\tilde{\alpha}_B = 0$).
Taking the divergence of the force balance equation, $\partial_i \sigma_{ij} = 0$ gives an equation for the isotropic strain
\begin{equation} \label{eq:iso-strain-Poisson}
    (B+\mu)\nabla^2 u_{kk} = - \tilde{\alpha} \, \partial_{i}\partial_{j}Q_{ij} \,,
\end{equation}
which for $\mathbf{Q} = \mathbf{Q}_{+\frac12}$ has the approximate solution $u^{+\frac12}_{kk}(\mathbf{r}) \approx -\frac{\tilde{\alpha}S_{1/2}(r)}{2(B+\mu)} \frac{x}{r}$ (see SI).
For contractile active stress ($\tilde{\alpha} > 0$), as found for instance in a tissue like a \emph{Hydra's} ectoderm \cite{West1978}, this strain field is tensile in front of the defect and compressive behind it [Fig.~\ref{fig:1}(b)].
In our model, this strain gradient leads to a free energy density gradient across the defect through the strain-ordering coupling $\chi_B^{}$.
Thus, remodeling of the nematic texture, Eq.~\eqref{eq:nematic-remodeling}, propels the defect in the direction that reduces this free energy. 
As the defect moves, the strain field generated by active stresses shifts along with it such that there will always be a strain gradient across the defect core, sustaining the effective force acting on the defect. 
This propulsion mechanism is markedly different from the fluid case, where the defect is advected by the velocity field it actively induces~\cite{giomi2013defect,pismen2013dynamics}. 

To estimate the defect propulsion force, we make the variational ansatz $\mathbf{Q}(\mathbf{r}, t) = \mathbf{Q}_{+\frac12}[\mathbf{r} - \mathbf{r}_0(t)]$, which yields (see Appendix~\ref{app:variational} for details) 
\begin{equation} 
    \gamma_{+\frac12} \partial_t \mathbf{r}_0 = 
    \boldsymbol{f}_{\chi_B^{}}^{+\frac12} + \boldsymbol{f}_{\chi}^{+\frac12} + \boldsymbol{f}_\mathrm{ext}, 
\end{equation}
with the effective friction (inverse mobility) $\gamma_{+\frac12} = \int\!\mathrm{d}r^2 (\nabla_jQ_{ik})^{2}$, and $\boldsymbol{f}_\mathrm{ext}$ accounting for any external effective forces acting on the defect (e.g.\ from interaction with other defects). The self-propulsion forces are given by
\begin{align}
    \boldsymbol{f}_{\chi_B^{}}^{+\frac12} &= -\chi_B^{} \int\! \mathrm{d} r^{2} \, \nabla u_{kk}^{+\frac12}(\mathbf{r}) \left[1 - S_{1/2}^{2}(r)\right],
    \\
    \boldsymbol{f}_{\chi}^{+\frac12} &= -\chi \int \! \mathrm{d}^2r \, \tilde{u}^{+\frac12}_{ij}(\mathbf{r}) \nabla Q^{+\frac12}_{ij}(\mathbf{r}).
\end{align}
In the SI, we show that $\boldsymbol{f}^{+\frac12}_{\chi}$ is independent of $\tilde{\alpha}$ and proportional to $\tilde{\alpha}_B$, i.e.\ it vanishes for $\tilde{\alpha}_B = 0$.
The factor $(S^{2}_{1/2} - 1)$ in $\boldsymbol{f}^{+\frac12}_{\chi_B}$ is nonzero only inside the defect core ($r \lesssim \sqrt{K}$), in agreement with the intuition that the strain gradient across the defect core drives defect propulsion.
For and isolated defect ($\boldsymbol{f}_\mathrm{ext} = 0$) and $\tilde\alpha_B=0$, numerical integration yields 
\begin{equation} \label{eq:f-chi_B}
    \mathbf{v}_0=\frac{\boldsymbol{f}_{\chi_B^{}}^{+\frac12}}{\gamma_{+\frac12}} \approx 0.4 \, \frac{\tilde{\alpha} \chi_B^{}\sqrt{K}}{\gamma_{+\frac12}(B+\mu)} \, \hat{\mathbf{p}}\;.
\end{equation}
Importantly, the direction of propulsion of the $+1/2$ defect is determined by both the sign of the active stress $\alpha$ and that of the coupling parameter $\chi_B^{}$. 
In particular, this allows a system with contractile activity to have a defect move towards its head when $\chi_B^{}$ is negative, i.e.\ when extensile strain reduces nematic order.
This is in contrast to active fluids, where the sign of activity alone controls the direction of propulsion such that $+1/2$ defects move towards their tail for contractile activity.

To numerically validate the analytic estimate for $\boldsymbol{f}_{\chi_B^{}}^{+\frac12}$, we simulate a $+1/2$ defect in a disk with suitable anchoring boundary conditions. Boundary anchoring leads to an effective force $\boldsymbol{f}_\mathrm{ext} = -\frac{\pi}{2} K \mathbf{r}_0/(R^2 - r_0^2)$ that pushes the defect towards the disk center and can be calculated by placing a $+1/2$ image charge at $(R^2/r_0^2) \mathbf{r}_0$. 
Equation \eqref{eq:f-chi_B} gives a good prediction for the position where the defect stalls ($\boldsymbol{f}_{\chi_B^{}}^{+\frac12} + \boldsymbol{f}_\mathrm{ext} = 0$); see Fig.~\ref{fig:1}(c).
Deviations are due to the variational ansatz that neglects distortions of the nematic texture \cite{Montagna.etal2025} and to approximations made when estimating the elastic strain.

In the complementary case of purely isotropic induced stress ($\tilde{\alpha} = 0$, $\tilde{\alpha}_B \neq 0$), the 
isotropic strain is azimuthally symmetric and, hence,  $\boldsymbol{f}_{\chi}^{+\frac12}$ vanishes.
Due to azimuthal symmetry, the deviatoric strain induced by isotropic stress has principal axes pointing along the radial and azimuthal directions. This implies that the strain alignment contribution  $\chi \tilde{u}_{ij} Q_{ij}$ in the free energy has opposite signs in the head and the tail of the defect [Fig.~\ref{fig:strain-aligment-force}(a)].
The resulting defect-propulsion force $\boldsymbol{f}_{\chi}^{+\frac12} \approx -\frac{\pi}{16} \frac{\tilde{\alpha}_B \chi \sqrt{K}}{B + \mu} \hat{\mathbf{p}}$ is derived in Appendix~\ref{app:isotropic-stress-propulsion}.

\begin{figure}
    \centering
    \includegraphics{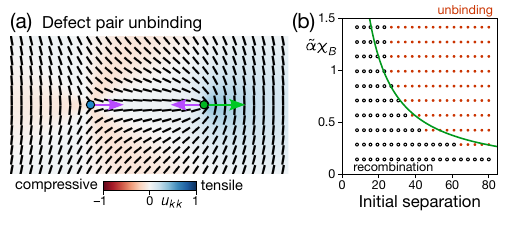}
    \caption{(a)~Defect pairs unbind when the propulsion force (green arrow) overcomes Coulomb attraction (purple arrows). The snapshot shows a cropped region from a larger simulation domain with free boundary conditions; see Movie~2.
    (b)~Analytic prediction (green line) vs numerical simulations testing defect unbinding.
    (Parameters: $\chi = 0$, $L_x = 500, L_y = 200$. Free boundary conditions: $\sigma_{ri}|_{r = R} = 0$.)
    }
    \label{fig:unbinding}
\end{figure}

\textit{Defect-pair unbinding.}
In a passive nematic liquid crystal, defects of opposite charge attract through a Coulomb-like force with strength $\pi K/(2d)$~\cite{chaikin1995principles}, where $d$ is the defect separation.
We therefore expect that a $+1/2$ defect that is oriented away from the $-1/2$ defect can escape its attraction beyond the critical separation
$d^* = -\frac{\pi \sqrt{K} \, (B + \mu)}{0.4 \, \tilde{\alpha} \chi_B^{}}\sim K/|\mathbf{v}_0|$, analogously to active fluids~\cite{giomi2013defect,shankar2018defect}.
To test this prediction, we simulated $\pm 1/2$ defect pairs in a rectangular geometry [Fig.~\ref{fig:unbinding}(a); Movie~2].
Indeed for strong enough active stress $\alpha$ and large enough initial separation $d$, the defect separation increases with time and the $+1/2$ defect escapes the $-1/2$ [Fig.~\ref{fig:unbinding}(b)].

\textit{Interaction of $+1/2$ defects in a confined domain.}
Biological tissue is often geometrically confined which can enforce the presence of topological defects through boundary anchoring \cite{Guillamat.etal2022} or through topological constraints on surfaces with non-zero Euler characteristic (e.g.\ spheroids \cite{Maroudas-Sacks.etal2021}).
To study the interaction of two $+1/2$ defects in a confined geometry, we consider a disk-shaped domain with perpendicular nematic anchoring along the boundary enforcing a net $+1$ topological charge.

\begin{figure}[t]
    \centering
    \includegraphics[width=\linewidth]{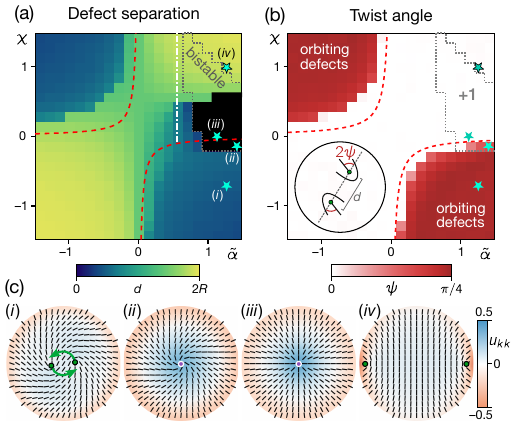}
    \caption{
    (a,b) Phase diagrams quantifying the steady-state defect separation (a) and ``twist angle'' (b) for two $+1/2$ defects in a disk geometry with normal boundary anchoring of the nematic.
    The regions in the top left and bottom right corner where $\psi > 0$ and $d > 0$, correspond to orbiting defect pairs.
    The dotted gray lines in (a,b) bound the region where $+1$ defects are stable (see SI). The white dot-dashed line in (a) shows the analytically predicted stability threshold [see Eq.~\eqref{eq:+1-threshold}].
    In the region which is black in (a), initially separated $+1/2$ defects merge to form a $+1$ defect. Above this is a bistable region, where $+1/2$ defects either repel or merge into a $+1$ depending on their initial distance (see SI).
    The red dashed lines show the linear stability prediction for spiral formation (see SI).
    (c)~Representative snapshots for parameter combinations labeled (\textit{i--iv}) in the phase diagram (see Movie~3). 
    In the spiral regime, $+1/2$ defects orbit around the origin [indicated by green arrows in (\textit{i})].
    (Domain radius $R = 16$; free boundary conditions: $\sigma_{ri}|_{r = R} = 0$.)
    }
    \label{fig:phase-diagram}
\end{figure}

Figures~\ref{fig:phase-diagram}(a) and \ref{fig:phase-diagram}(b) show the numerical phase diagram in the $(\tilde{\alpha}, \chi)$-plane where we characterize the configurations in terms of the defect separation $d$ and the twist angle $\psi$ measuring the orientation of the $+1/2$ polarization (head--tail axis) relative to the radial direction. 
Movie~3 and representative snapshots in Fig.~\ref{fig:phase-diagram}(c) show the four distinct types of behaviors found in simulations:
(\textit{i}) a twisted nematic texture that spontaneously breaks chiral symmetry with two orbiting $+1/2$ defects;
a stable $+1$ defect in the center of the domain (\textit{ii}, \textit{iii}); and (\textit{iv}) a configuration where the $+1$ breaks into two $+1/2$s pushed to the boundary.
For large $\chi$, we observe bistability (hysteresis) where the steady state configuration depends on the initial condition (see SI).

The transition to twisted nematic textures for $\tilde{\alpha} \chi < 0$ can be understood in terms of geometric incompatibility: For a $+1$ defect with phase $\psi$ (the director angle relative to the polar angle from the defect core) the nematogen-induced stress generates intrinsic Gaussian curvature of magnitude $\sim \cos(2\psi)$ concentrated at the core~\cite{Mostajeran2015,Duffy.Biggins2020} (see Appendix~\ref{app:core-melting}). Therefore, in a flat geometry, there will be residual stresses unless $\psi = \pm\pi/4$.
Starting from an achiral state $\psi = 0, \pi/2$, relaxation of elastic energy will spontaneously break chiral symmetry to $\psi = \pm\pi/4$ within a boundary layer of width $\sim \sqrt{K}$. 
A linear stability analysis taking into account this boundary layer [see SI; dashed red lines in Fig.~\ref{fig:phase-diagram}(a,b)] yields good agreement with simulations.
In a large part of the parameter space, the $+1$ charge is resolved into a pair of $+1/2$ defects. Due to the spiral-like texture, the defects' head-tail axes are at an angle to the radial direction, such that their self-propulsion drives cyclic motion [Fig.~\ref{fig:phase-diagram}(c-i)].

A $+1$ defect forms in the center of the disk when the inward pointing radial component of the defects' self-propulsion force is strong enough to push them closer than the core radius. The active stress required to stabilize a $+1$ defect in the center can be estimated from energetic arguments (Appendix~\ref{app:core-melting}), yielding good agreement with numerics [Fig.~\ref{fig:phase-diagram}(a)].
Notably, this ``active stabilization'' becomes more effective in large domains, in contrast to passive $+1$ defects that are stabilized by confinement of size  similar to the nematic coherence length~\cite{Manyuhina.etal2015}. Alternative mechanisms for active stabilization of $+1$ defects have recently been studied in models with shape-tension coupling~\cite{Dye.etal2021,Rozman.etal2023}.

\textit{Role of substrate traction.}
Tissues studied \textit{in vitro} are often attached to a rigid substrate \cite{Guillamat.etal2022,Kawaguchi.etal2017,saw2017topological} which can be incorporated by introducing a spring-like traction force $-k u_i$ so that force balance reads $\partial_j\sigma_{ij}=ku_i$. Long-ranged elastic interactions then become screened, with a screening length $\ell_\mathrm{el} \sim \sqrt{(B+\mu)/k}$, analogously to viscous screening in active fluids~\cite{ronning2022flow}. Screening eliminates system-size dependence arising form the long-range nature of elastic forces (e.g., in the stability condition for $+1$ defects Eq.~\eqref{eq:+1-threshold}, see Appendix~\ref{app:core-melting}).

Importantly, traction provides an additional mechanism for feedback between mechanics and nematic order, as it allows a term $\lambda u_i u_j Q_{ij}$ in the free energy Eq.~\eqref{eq:free-energy} describing alignment of $Q_{ij}$ to the displacement field. This may 
represent, for instance, alignment of actin fibers along the direction of traction force. In the limit of strong coupling to the substrate, we can neglect elastic stresses and solve force balance for the displacement field as $u_i \approx \frac{\tilde{\alpha}}{k}\partial_j Q_{ij}$ to leading order.
Alignment with displacement then gives a contribution $(\lambda \tilde{\alpha}^2/k) \partial_k Q_{ik} \partial_l Q_{jl}$ to the dynamics of $Q_{ij}$. 
For parallel alignment of nematic order and displacement/traction, this causes $+1/2$ defects to
move towards their head, independent of the sign of activity. In a purely passive system, the defect propulsion through this mechanism will be transient until distortion of the nematic texture leads to arrest of motion, as has recently been shown in nematic fluids \cite{Montagna.etal2025}.

\textit{Discussion.}
In active nematics $+1/2$, defects are generically expected to self-propel due to their polar symmetry. Such motility is transient in equilibrium, but becomes sustained in active systems. It has been studied extensively in \emph{fluids} where defects are advected by the active flows they generate. Here, we have proposed new mechanisms for self-sustained defect propulsion in active nematic \emph{solids} where defects move relative to the elastic medium via local remodeling of the nematic texture.

These mechanisms may provide an explanation for a puzzling observation in cell monolayers, where defect motion often suggests extensile activity despite contractile forces at the cell scale~\cite{saw2017topological,Kawaguchi.etal2017,balasubramaniam2021investigating}. Several explanations have been proposed~\cite{vafa2021fluctuations,killeen2022polar,bonn2022fluctuation,dedenon2024noise}, including most recently that cell motion is primarily driven by propulsive traction with the substrate and that the nematic texture of cell shapes and associated topological defects is a passive response~\cite{nejad2024stress,Bera.etal2025}. The mechanism proposed here provides an alternative explanation that does not require cell motility. The telltale sign of this mechanism is that defects move relative to the cells, as it is observed in regenerating \textit{Hydra}~\cite{Maroudas-Sacks.etal2021}. It will be interesting to investigate this in other tissues via the simultaneous tracking of topological defects and individual cells.

In tissues, coupling between stress and nematic order might be mediated by intermediary biochemical processes and signaling molecules (morphogens). Coupling to morphogen concentration gradients provides additional mechanisms for defect unbinding and propulsion~\cite{Wang.etal2020}. An important direction for future work is therefore to augment the current model by including such morphogen concentration fields which can, in turn, be sourced by the stress focused near topological defects \cite{Ferenc.etal2020,Maroudas-Sacks.etal2024confinement}.
Such a feedback loop may provide an alternate mechanism for the stabilization of the $+1$ defect at the head of \textit{Hydra}~\cite{Wang.etal2020,Maroudas-Sacks.etal2024confinement}.


Finally, in-plane stress can also drive out-of-plane deformations, generating 3D shapes \cite{Modes.etal2011}, as has recently been utilized for tissue engineering \cite{Guillamat.etal2025}. Investigating the  interplay of dynamic topological defects and 3D shape change is an important future direction with applications from morphogenesis and tissue engineering to soft robotics, where this interplay might guide the design of self-regulated actuators \cite{Zeng.etal2017}, pumps \cite{Lee.Bhattacharya2021} or locomotion mechanisms \cite{Maghsoodi.Bhattacharya2022,SeungChoi.SeokKim2024,Guo.etal2025}.

\vspace*{1em}
We thank Kinneret Keren, Ananyo Maitra, Carl Modes, Michael Moshe, and Marko Popovi\'c for insightful discussions and the Kavli Institute for Theoretical Physics for hospitality during the final stages of completion of this work. FB acknowledges support by the Gordon and Betty Moore Foundation post-doctoral fellowship (grant \#2919). MO acknowledges support from the University of California Santa Barbara College of Creative Studies’ Francesc Roig Summer Undergraduate Research Fund. MCM was supported by the National Science Foundation award DMR-2041459.

FB and MO contributed equally to this work.

\bibliography{ActiveSolid.bib}

\clearpage

\appendix

\section{Connection to metric elasticity and shape programming}
\label{app:metric-elasticity}

Our model has important connections to activated nematic elastomers which are often modeled using the framework of metric elasticity. There, the internal conformation change of the elastomer is described through a ``target strain'' that deforms the internal reference configuration of the material~\cite{griniasty2019curved,Duffy.Biggins2020,warner2020topographic}.
In the free energy Eq.~\eqref{eq:free-energy}, these target strains are captured by the elasto-nematic couplings $\chi_B^{}$ and $\chi$. Indeed, these terms can be rewritten as target strains by completing the square \cite{Maitra.Ramaswamy2019}.
When the target strains are geometrically incompatible with a flat realized configuration, i.e.\ when they have have intrinsic curvature $\partial_i \partial_j Q_{ij} \neq 0$, residual elastic stresses remain even after the elastic energy has been minimized. When the elastic sheet is allowed to deform out of plane, the residual stresses lead to buckling. This is the mechanism used for shape programming in nematic elastomers~\cite{griniasty2019curved,Duffy.Biggins2020}.
In the language of geometric charges, $\tilde{\alpha} Q_{ij}$ plays the role of a curvature quadrupole~\cite{Lemaitre.etal2021}.

\section{Variational ansatz for defect propulsion}
\label{app:variational}

We use a variational formulation to find the dynamics of a $+1/2$ defect due to elasto-nematic coupling.
To this end, we introduce the Rayleigh dissipation functional
\begin{equation}
    \mathcal{R}\bigl[\dot{\mathbf{Q}}\bigr] = \frac{\gamma}{2} \int\!\mathrm{d}^2r \left(\partial_t Q_{ij}\right)^2,
\end{equation}
such that we can write the $\mathbf{Q}$ dynamics Eq.~\eqref{eq:nematic-remodeling} in a variational form
\begin{equation}
    \gamma \partial_t Q_{ij} = \frac{\delta \mathcal{R}}{\delta (\partial_t Q_{ij})} = -\frac{\delta \mathcal{F}}{\delta Q_{ij}} \, .
\end{equation}
Physically, this equation expresses the power balance between the change of free energy and dissipation. 
To find the approximate equation of motion for the $+1/2$ defect, we substitute the ansatz $\mathbf{Q}(\mathbf{r}, t) \approx \mathbf{Q}^{+\frac12}[\mathbf{r} - \mathbf{r}_0(t)]$ into the variational equation, to obtain
\begin{align}
    \frac{\delta \mathcal{R}}{\delta \dot{\mathbf{r}}_0} &= -\frac{\delta \mathcal{F}}{\delta \mathbf{r}_0} \,,\\
    \int \!\mathrm{d}^2r \, (\dot{\mathbf{r}}_0 \cdot \nabla Q_{ij}^{+\frac12}) \nabla Q_{ij}^{+\frac12} &=
    \int\!\mathrm{d}^2r \left.\frac{\delta\mathcal{F}}{\delta Q_{ij}} \right|_{\mathbf{Q}^{+\frac12}} \! \nabla Q^{+\frac12}_{ij} \,.
    \label{eq:power-balance-integral}
\end{align}
Evaluating the left-hand side yields the effective friction coefficient
\begin{multline}
    \int_0^{R} \! \mathrm{d}r \int_{0}^{2\pi} \! r\mathrm{d}\theta \left[\left(S_{1/2}'\right)^{2}\hat{r}_i\hat{r}_j + \frac{S_{1/2}^2}{r^2} \hat{\theta}_i\hat{\theta}_j\right] \\
    =
    \frac{1}{2} \delta_{ij} \int_{0}^{R}\! \mathrm{d}r \left[\frac{S_{1/2}^{2}}{r} + r(S_{1/2}')^{2}\right]
    =: \gamma^{+\frac12} \delta_{ij} \;,
\end{multline}
The effective friction (inverse mobility) of a defect diverges logarithmically with increasing system size because defect motion requires global remodeling of the nematic texture. 
In a system with multiple defects, the dissipation rate depends on the locations of the other defects and is ``screened'' on the scale of the typical defect separation~\cite{denniston1996disclination,toth2002hydrodynamics}.

To evaluate the right-hand side of Eq.~\eqref{eq:power-balance-integral}, we use integration by parts and the fact that $\mathbf{Q}^{+\frac12}$ remains a free energy minimum of $\mathcal{F}_Q^{(0)}$ under translation. Therefore, only the terms coupling $\mathbf{Q}$ to strain remain:
\begin{multline} \label{eq:defect-force}
    \gamma^{+\frac12} \partial_t \mathbf{r}_0 = \boldsymbol{f}^{+\frac12} := 
    \underbrace{{}-\chi_B^{} \int\!\mathrm{d}^{2}r \, \left[1 - S_{1/2}^{2}(r) \right] \nabla u_{kk}(\mathbf{r})}_{\displaystyle \boldsymbol{f}^{+\frac12}_{\chi_B^{}}}
    \\
    \underbrace{{}-\chi \int\!\mathrm{d}^{2}r \, \tilde{u}_{ij}(\mathbf{r}) \nabla Q_{ij}^{+\frac12}(\mathbf{r})}_{\displaystyle \boldsymbol{f}^{+\frac12}_{\chi}} .
\end{multline}

Alternatively, this equation can be obtained through a systematic perturbation ansatz as in Ref.~\cite{shankar2018defect}.
First, rewrite Eq.~\eqref{eq:nematic-remodeling} as
\begin{equation}
    \gamma \partial_t Q_{ij} = F^{(0)}_{ij} + F^{(1)}_{ij} \;,
\end{equation}
where $F^{(0)}_{ij}$ and $F^{(1)}_{ij}$ collect terms to zeroth and first order in $\chi, \chi_B^{}$.
Using the fact that $\nabla Q_{ij}^{+\frac12}$ is in the nullspace of the linearization of $F^{(0)}_{ij}$ yields the solvability condition 
\begin{align}
    \int\!\mathrm{d}^2r \left[F^{(1)}_{ij} - \dot{\mathbf{r}}_0 \cdot \nabla Q_{ij}^{+\frac12} \right] \nabla Q^{+\frac12}_{ij} = 0 \,,
\end{align}
which recovers the equation of motion of the defect position Eq.~\eqref{eq:defect-force} we obtained using the variational ansatz.

\section{Strain-alignment mediated defect propulsion}
\label{app:isotropic-stress-propulsion}

\begin{figure}
    \centering
    \includegraphics{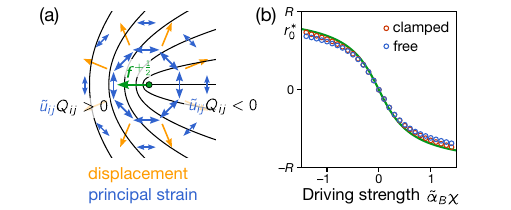}
    \caption{
    (a)~Illustration of the mechanism for defect propulsion due to active bulk stress and strain alignment. Contractile active bulk stress induces a displacement field (orange arrows) pointing radially away from the defect. The resulting strain (blue double arrows) is oriented azimuthally around the defect, implying that the strain-alignment energy density $\chi u_{ij} Q_{ij}$ has opposite signs in the defect's head and tail, causing an effective self-propulsion force towards the defects head for $\chi > 0$ (green arrow).
    (b)~The steady state defect position due to a balance of defect self-propulsion and Coulomb repulsion from the boundary in a disk-shaped domain predicted from Eq.~\eqref{eq:propulsion-strain-alignment} agrees well with numerics. The mechanical boundary conditions (clamped: red circles; free: blue circles) plays only a minor role.
    (Parameters: $R=16, \chi = 2, \chi_B^{} = 0$.)
    }
    \label{fig:strain-aligment-force}
\end{figure}

Strain alignment drives defect an propulsion force $\boldsymbol{f}_\chi^{+\frac12}$ if the alignment energy density $\chi \tilde{u}_{ij} Q_{ij}$.
To calculate this force, we consider the contributions from deviatoric active stress ($\sim \tilde{\alpha}$) and isotropic active stress ($\sim \tilde{\alpha}_B$) separately. 
In the SI, we show that the former contribution vanishes identically.
To find the contribution due to isotropic active stress, first observe that the stress profile is azimuthally symmetric around the defect core. Therefore the induced elastic strain will have the same symmetry and its principal axes will be radial and azimuthal as illustrated in Fig.~\ref{fig:strain-aligment-force}(a).
We use the Airy stress formalism (see SI) and the approximation $S_{1/2}(r) \approx \Theta(r - \sqrt{K})$, to find the deviatoric strain
\begin{equation}
    \tilde{u}^{+\frac12}_{rr} \approx -\frac{\tilde{\alpha}_B K}{8 (B + \mu) \, r^2} \, \Theta(r - \sqrt{K}) \,.
\end{equation}
Substituting into the expression for $\boldsymbol{f}^{+\frac12}_\chi$ [cf.\ Eq.~\ref{eq:defect-force}] yields
\begin{align}
    \boldsymbol{f}^{+\frac12}_\chi &= -\frac{\pi \chi}{2} \hat{\mathbf{p}} \, 
    \int_0^R \!\mathrm{d}r \, r^2 \, \tilde{u}^{+\frac12}_{rr}(r) \, \partial_r \!\left(\frac{S_{1/2}(r)}{r}\right) \,, \\ 
    &\approx -\frac{\pi}{16} \frac{\tilde{\alpha}_B \chi \sqrt{K}}{B + \mu} \hat{\mathbf{p}} \,.  \label{eq:propulsion-strain-alignment}
\end{align}
This result agrees well with numerics [ Fig.~\ref{fig:strain-aligment-force}(b)].

Taken together, we have
\begin{equation}
    \boldsymbol{f}^{+\frac12} \approx  \left( 0.4 \, \tilde{\alpha} \chi_B^{} - \frac{\pi}{16}\tilde{\alpha}_B \chi \right) \frac{\sqrt{K}}{B + \mu} \, \hat{\mathbf{p}} \, ,
\end{equation}
showing the ``duality'' of the two defect propulsion mechanisms: the first operating through the combination of deviatoric nematogen-induced stress $(\sim \tilde{\alpha})$ and strain ordering $(\sim \chi_B^{})$; the second operating through the combination isotropic nematogen-induced stress $(\sim \tilde{\alpha}_B)$ and strain alignment $(\sim \chi)$.

Note that in the purely passive case $\alpha = \alpha_B = 0$ (i.e.\ $\tilde{\alpha} = -\chi$, $\tilde{\alpha}_B = -\chi_B^{}$), the propulsion force for a defect at the center of a disk-shaped domain does not vanish: 
\begin{equation} \label{eq:propulsion-passive}
    \boldsymbol{f}^{+\frac12}_\mathrm{passive} \approx 0.2 \, \frac{\chi \chi_B^{} \sqrt{K}}{B + \mu} \, \hat{\mathbf{p}} \, .
\end{equation}
In this case the defect moves away from the domain center by extracting elastic energy from the elastic medium. Of course, this passive motion is transient and ends when the free energy $\mathcal{F}$ has reached a minimum.
Calculating this minimum-energy position analytically is cumbersome because one can no longer exploit azimuthal symmetry.

\section{+1 stability criterion}
\label{app:core-melting}

The isotropic strain concentrated at the defect core is sufficiently strong to completely ``melt'' the nematic order there via isotropic elasto-nematic coupling. 
In this case it is energetically favorable to localize the entire topological charge in the melted region. 
We can therefore estimate the active stress required to stabilize a $+1$ defect by requiring that the core is in the isotropic regime, i.e.\ $\chi_B^{} u_{kk}(r = \sqrt{K}) < -1$.

There is a family of $+1$ defects parametrized by the ``phase'' $\psi_0$
\begin{equation}
    \mathbf{Q}^{+1}(\mathbf{r}; \psi_0) = \frac12 S_1(r)
    \begin{pmatrix}
        \cos(2\psi_0 + 2\theta) & \sin(2\psi_0 + 2\theta) \\
        \sin(2\psi_0 + 2\theta) & -\cos(2\psi_0 + 2\theta)
    \end{pmatrix},
\end{equation}
where $\psi_0 = 0, \pi/2$ correspond to an aster and vortex respectively. For other values of $\psi_0$, the texture is a spiral with pitch angle $\psi_0$.
Substituting into Eq.~\eqref{eq:iso-strain-Poisson} yields
\begin{align}
    (B + \mu)\nabla^2 u^{+1}_{kk} \approx -2\pi\tilde{\alpha} \cos(2\psi_0) \, \delta^2(\mathbf{r})\,,
\end{align}
where we have made a far-field approximation (monopole term in multipole expansion) for $\partial_i \partial_j Q^{+1}_{ij}$ (see SI for details).
Note that the induced elastic strain vanishes for the $\psi_0 = \pi/4$ spiral. In this case, the ``target strain'' induced by active stress and strain-alignment is geometrically compatible, i.e.\ has no intrinsic curvature. 

This is solved by the Green's function for the Laplacian
\begin{equation}
    (B + \mu) u^{+1}_{kk}(r) = -\tilde{\alpha} \cos(2\psi_0) \log \frac{r}{R}\,,
\end{equation}
where we assume that isotropic strain (and hence pressure) vanish at the boundary. 
Substituting into $\chi_B^{} u_{kk}(r = \sqrt{K}) < -1$ gives the estimated stability condition for $+1$ defects 
\begin{equation} \label{eq:+1-threshold}
   \frac{\tilde{\alpha}\chi_B^{}}{\mu + B} \cos(2\psi) < -\frac{1}{\log(R/\sqrt{K})}\,.
\end{equation}
Comparing to numerics shows that Eq.~\eqref{eq:+1-threshold} slightly underestimates the magnitude of $\tilde{\alpha}$ required to stabilize the $+1$ defect.
Nonetheless, Eq.~\eqref{eq:+1-threshold}, provides several qualitative insights.
First, in the region where the texture spontaneously twists into a spiral $+1$ defects can only be stabilized when the spiral pitch angle $\psi_0$ is sufficiently small. Indeed, there is only a very small regime with stable $+1$ spirals near the onset of spiral formation [see (\textit{ii}) in Fig.~\ref{fig:phase-diagram}].
Second, a $+1$ defect stabilizes more readily for a larger domain, since more active stress builds up over a larger area. 
Thus, the defect stabilization through elasto-nematic coupling is very different from stabilization through spatial confinement alone, where the latter requires a small domain size on the order of the nematic coherence length $\sqrt{K}$.
In the presence of spring-like attachment to a substrate, elasticity becomes screened such that the system size in Eq.~\eqref{eq:+1-threshold} is replaced by the screening length (see SI).  

\end{document}